# WFIRST-2.4: What Every Astronomer Should Know

Science Definition Team
D. Spergel[1], N. Gehrels[2]
J. Breckinridge[3], M. Donahue[4], A. Dressler[5], B. S. Gaudi[6], T. Greene[7], O. Guyon[8], C. Hirata[3]
J. Kalirai[9], N. J. Kasdin[1], W. Moos[10], S. Perlmutter[11], M. Postman[9], B. Rauscher[2], J. Rhodes[12], Y. Wang[13]
D. Weinberg[6], J. Centrella[14], W. Traub[12]

Consultants
C. Baltay[15], J. Colbert[16], D. Bennett[17], A. Kiessling[12], B. Macintosh[18], J. Merten[12], M. Mortonson[6], M. Penny[6]
E. Rozo[19], D. Savransky[18], K. Stapelfeldt[2], Y. Zu[6]

Study Team
C. Baker[2], E. Cheng[20], D. Content[2], J. Dooley[12], M. Foote[12], R. Goullioud[12], K. Grady[2], C. Jackson[21], J. Kruk[2]
M. Levine[12], M. Melton[2], C. Peddie[2], J. Ruffa[2], S. Shaklan[12]

1 Princeton University
2 NASA/Goddard Space Flight Center
3 California Institute of Technology
4 Michigan State University
5 Carnegie Institution for Science
6 Ohio State University
7 NASA/Ames Research Center
8 University of Arizona
9 Space Telescope Science Institute
10 Johns Hopkins University
11 University of California Berkeley/Lawrence Berkeley National Laboratory
12 Jet Propulsion Laboratory/California Institute of Technology
13 University of Oklahoma
14 NASA Headquarters
15 Yale University
16 Infrared Processing and Analysis Center/California Institute of Technology
17 University of Notre Dame
18 Lawrence Livermore National Laboratory
19 Stanford Linear Accelerator Laboratory
20 Conceptual Analytics
21 Stinger Ghaffarian Technologies

# 1 INTRODUCTION

A Wide Field Infrared Space Telescope (WFIRST) was the top-ranked priority for a large space mission in the *New Worlds/New Horizons* (*NWNH*) report of the 2010 National Academy of Sciences Decadal Survey for Astronomy, *NWNH* recognized that recent improvements in infrared detector technology enabled a mission that would achieve dramatic advances across a wide range of astrophysics, including dark energy, the demographics of planetary systems, and studies of galaxy evolution, quasar evolution, and stellar populations of the Milky Way and its neighbors. The WFIRST Science Definition Team (SDT), building on several generations of work by previous teams, improved on the design originally submitted to the Decadal Survey and defined an observing program capable of addressing the broad goals laid out by *NWNH*. Their final report (Green et al. 2012) presented two Design Reference Missions: DRM1, with a 1.3m off-axis telescope and a 0.375 deg$^2$ field of view, and DRM2 with a 1.1m telescope, a 0.585 deg$^2$ field of view, and a shorter primary mission (3 years vs. 5 years). As that report was nearing completion, a pair of Hubble-quality, 2.4m telescopes, originally built for a different program in another agency, were offered to NASA, presenting an extraordinary new opportunity for space science.

In response to this opportunity, NASA commissioned a new SDT to create a design reference mission for WFIRST using one of these 2.4m telescopes. This short article describes the highlights of this DRM, hereafter referred to as WFIRST-2.4, and serves as an entry point to the much more detailed description in the SDT report (Spergel et al. 2013). The most important difference between WFIRST-2.4 and the 1.3m DRM1 is, of course, the larger aperture, which affords both greater sensitivity and sharper angular resolution. There are some losses – a ~30% reduction in field of view, and a 2.0μm wavelength cutoff (vs. 2.4μm) because of a higher telescope operating temperature – but for most investigations WFIRST-2.4 is a substantially more powerful facility than DRM1. The integral field unit (IFU) considered as an augmentation option in DRM1 has been incorporated into the baseline design of WFIRST-2.4, improving the performance of the supernova cosmology survey and adding a new science capability for other investigations. At NASA's direction, the SDT also considered the option of an on-axis coronagraphic instrument, which would take advantage of WFIRST-2.4's high angular resolution to achieve direct imaging and spectroscopy of nearby giant planets and debris disks.

Section 2 of this article summarizes the WFIRST-2.4 DRM, including the major components of the observing program. Section 3 summarizes the goals and anticipated performance in a variety of science areas, from planets to cosmology, including a list of the many ideas for Guest Observer programs that were submitted to the SDT. Section 4 offers some brief concluding remarks. For supporting details behind these summaries, we refer readers to the 190-page SDT report; figures and tables in this article are excerpted from that report.[1] Throughout this article, we provide brief comparisons between the capabilities of WFIRST-2.4 and those of the James Webb Space Telescope (JWST), the Large Synoptic Survey Telescope (LSST), and the European-led Euclid mission. These four ambitious projects are highly complementary, and for many investigations they will be much more powerful in combination than any one would be on its own.

# 2 THE WFIRST-2.4 DESIGN REFERENCE MISSION

The basis of WFIRST-2.4 is the optical telescope assembly (OTA) made available to NASA, comprised of a 2.4-m primary mirror, a secondary mirror, and associated support and control structures. A tertiary mirror provides a three-mirror anastigmat optical configuration that produces diffraction-limited images over a wide field of view. The spacecraft design is based on that of the currently operating Solar Dynamics Observatory. After considering several options, the SDT chose a geosynchronous orbit with 28.5-degree inclination, motivated largely by the high data rate achievable from geosynchronous orbit and secondarily by the potential for robotic servicing. WFIRST-2.4 is designed to be robotically serviceable, should a future robotic servicing capability be deployed in geosynchronous orbit. The DRM adopts a telescope operating temperature of 270K, with the near-IR focal plane array at approximately 118K. This temperature degrades sensitivity beyond 2 microns, and the wavelength cutoff for the near-IR instruments is set at 2.0 μm. Future work can investigate the possibility of lower operating temperatures and extension of the wavelength range. Compared to DRM1, the effective area of the WFIRST-2.4 point spread function (PSF) is smaller by a factor of 1.9. This is less than the naively expected ratio of the primary mirror areas because of the on-axis vs. unobstructed optics, but it still represents an important gain for many scientific programs, leading to a higher source density for weak lensing studies, improved deblending for microlensing and stellar population studies, greater sensitivity to point sources and compact galaxies, and higher astrometric accuracy.

---

[1] Radical compression necessarily entails simplification, and the SDT report remains the definitive reference on the WFIRST-2.4 DRM. It also provides extensive references to background literature, which we have deliberately kept minimal in this brief summary.

Table 1 summarizes the key characteristics of the WFIRST-2.4 instruments and observing program. The wide-field imager, illustrated in Figure 1, is comprised of 18 H4RG-10 HgCdTe detectors in a 6×3 rectangular array. The pixel scale is 0.11 arcsec, and there are more than 300 Mpixels covering an active area of 0.281 deg$^2$. This camera is fed by a filter wheel containing one wide filter for the microlensing survey, five broad-band filters spanning the wavelength range 0.76-2.0 μm, and a grism to provide resolution R ≈ 700 slitless spectroscopy for the galaxy redshift survey. The WFIRST-2.4 field is ~200 times bigger than the IR channel of HST's WFC3, which has already been a tremendously successful scientific tool. The wide field of WFIRST-2.4 is a powerful complement to the large aperture and still higher angular resolution of JWST; for example, by surveying thousands of deg$^2$ WFIRST-2.4 can find rare, luminous objects at very high redshifts, which can be studied in detail with JWST. The other component of the wide-field instrument is an IFU spectrometer, with a 3.0" × 3.15" field, spectral resolution R ≈ 100, and wavelength range 0.6-2.0 μm. The IFU will be used for spectroscopy and spectrophotometry of supernovae in the dark energy program, and it is a resource for measuring redshifts of faint galaxies and obtaining spatially resolved spectroscopy of medium redshift galaxies.

An optional second instrument is a coronagraphic imager and spectrograph, capable of directly detecting companions to nearby bright stars down to contrast ratios of 10$^{-9}$ and angular separations between 0.2 and 2 arcseconds. The integral field spectrograph is capable of acquiring spectra of these companions with resolutions of R~70 over the full wavelength range 400-1000 nm. Equipped with such an instrument, WFIRST-2.4 would be able to survey hundreds of nearby stars, enabling the characterization of dozens of known cool Jupiter-mass companions, the discovery and characterization of a similar number of cool Jupiter and Neptune companions, and the detection and characterization of debris disks in systems containing a few times the solar system's level of dust.

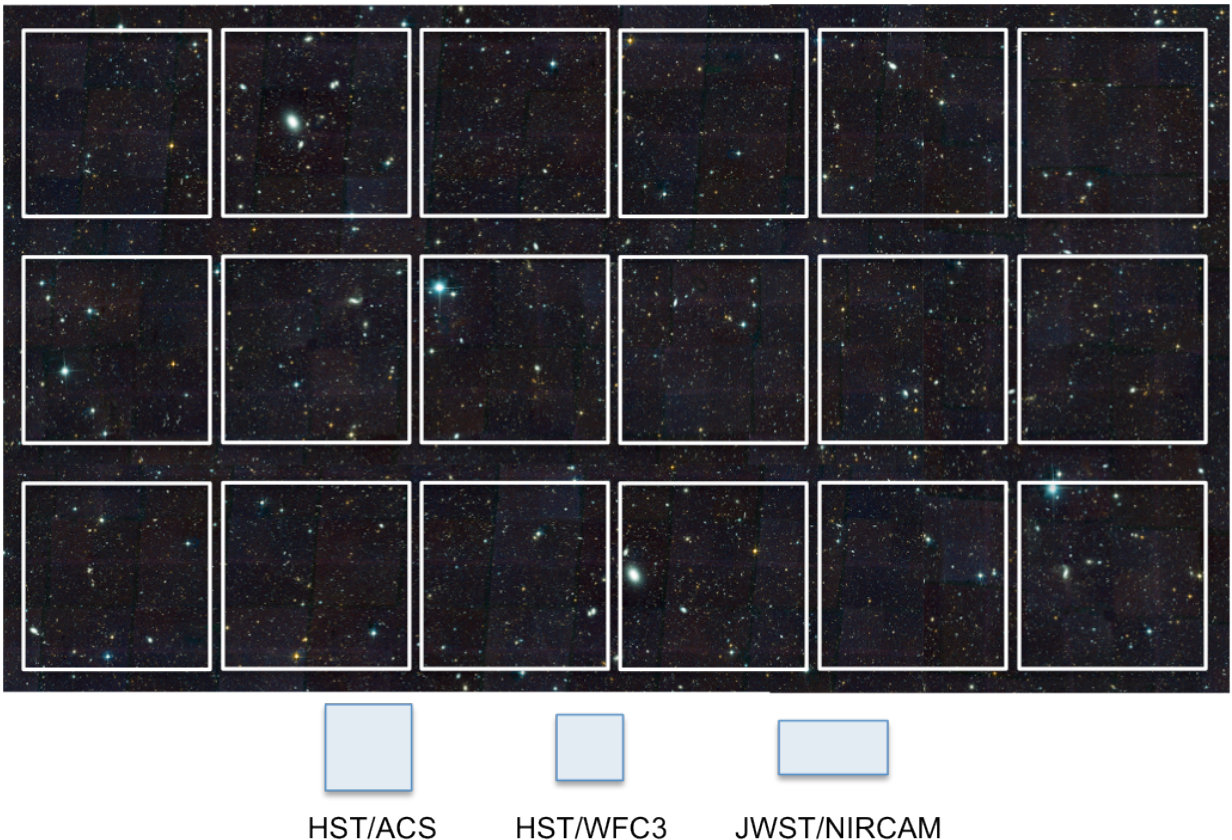

Figure 1: Field of view of the WFIRST-2.4 wide field instrument. Each square is a 4k x 4k HgCdTe sensor array, with pixels mapped to 0.11 arcseconds on the sky. The field of view extent is about 0.79 x 0.43 degrees. This is ~200 times the area of the IR channel on HST's WFC3, which is shown to scale, along with the fields of the HST ACS and JWST NIRCam.

To present definite forecasts for the scientific performance of WFIRST-2.4, we have adopted a notional observing program for a 5-year prime mission, which is summarized in Table 1 and detailed in subsequent sections. WFIRST-2.4 will support a wide range of science programs during its primary mission. Each of these programs has unique constraints involving the field of regard, cadence, and S/C roll angles. Moreover, observations in GEO are subject to viewing constraints on daily (motion around the Earth), monthly (Moon avoidance), yearly (Sun aspect angle), and secular (orbital precession) timescales. The SDT therefore constructed an existence proof of a possible observing plan, which showed that the strategic science programs are all mutually compatible, while simultaneously enabling a robust GO program. The actual allocation of observing time will be decided closer to the launch of WFIRST-2.4, based on scientific and technical developments between now and then. If the optional coronagraph is adopted, we assume that the prime mission will be extended to 6 years, enabling a robust coronagraph science program without reducing the reach of other programs.

The WFIRST-2.4 observing program has four main components, to be carried out over five years. The microlensing survey will monitor 2.81 deg$^2$ of the Galactic bulge, with 15-minute cadence, over six 72-day seasons, detecting thousands of exoplanets via the perturbations they produce on microlensing light curves. The high-latitude survey will carry out imaging and spectroscopy of a 2000 deg$^2$ area over 1.9 years of observing time, providing weak lensing shape measurements of 500 million galaxies and emission line redshifts of more than 20 million galaxies at redshifts $1 < z < 3$. The supernova survey will use 0.5 years of observing time over a 2-year period, with three tiers of imaging to discover supernovae and measure their light curves, and IFU spectroscopy of more than 2700 Type Ia SNe at $0.2 < z < 1.7$ to measure redshifts, spectral diagnostics, and well calibrated fluxes in synthetic filter bands matched across redshift. A Guest Observer program is allocated 1.4 years of observing time, spread across the 5-year mission. If the coronagraph is incorporated, the prime mission would be expanded to six years, with one year of coronagraphy observations interspersed with other programs.

Figure 2 shows the imaging depth achieved by the WFIRST-2.4 high-latitude survey (HLS), in comparison to LSST (after 10 years of operation) and Euclid. In an AB-magnitude sense, WFIRST-2.4 imaging is well matched to the i-band depth of LSST. The AB magnitude limits for LSST in g and r are fainter; however, a typical z>1 LSST weak lensing source galaxy has r-J or i-H color of about 1.2, so even here the WFIRST-2.4 imaging depth remains well matched, and of course the angular resolution is much higher. The Euclid IR imaging is ~2.5 magnitudes shallower than WFIRST-2.4 (and severely undersampled by its 0.3" near-IR pixels). The Euclid weak lensing survey uses a wide optical filter and reaches an AB-magnitude depth similar to what LSST achieves in each of its g, r, and i filters. The cumulative imaging depth in the WFIRST-2.4 supernova survey fields, which will be observed many times over a 2-year interval, reaches 0.5 – 2.5 magnitudes fainter than the HLS (see Table 1).

Figure 3 shows the emission line sensitivity of the HLS spectroscopic survey. For moderately extended sources, the 7$\sigma$ detection threshold is $\approx$ 1.0–1.5×10$^{-16}$ erg s$^{-1}$ cm$^{-2}$ depending on wavelength. The primary target is H$\alpha$ at $1.05 < z < 2$, but it is also possible to extend to higher redshift, $1.7 < z < 2.9$, using galaxies with strong [OIII] emission. Our forecasts based on recent estimates of the H$\alpha$ and [OIII] galaxy luminosity functions predict ~ 20 million H$\alpha$ galaxies, and ~ 2 million [OIII] galaxies at z > 2. At z = 1.5, the predicted comoving space density of the galaxy redshift survey is 1.5 × 10$^{-3}$ Mpc$^{-3}$. This is 16 times higher than the predicted space density for Euclid (from the same luminosity function), providing much better sampling of structure, over a smaller area of sky, for studies of baryon acoustic oscillations, redshift-space distortions, and higher order galaxy clustering. The "narrow/deep" strategy of WFIRST-2.4 nicely complements the "shallow/wide" strategy of Euclid.

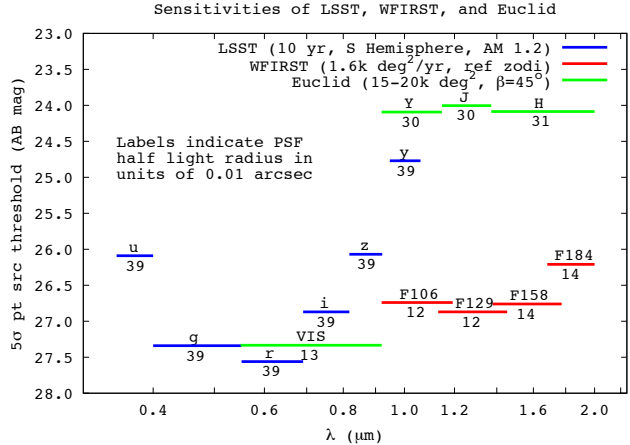

**Figure 2:** Depth of the WFIRST-2.4 high-latitude imaging survey (red), expressed in AB magnitudes for a 5$\sigma$ point source detection, compared to the expected depth of the Euclid (green) and LSST (blue) imaging surveys. Labels below each bar indicate the size of the PSF (specifically, the EE50 radius) in units of 0.01 arcsec. The near-IR depth of the the WFIRST-2.4 HLS is well matched to the optical depth of LSST (10-year co-add).

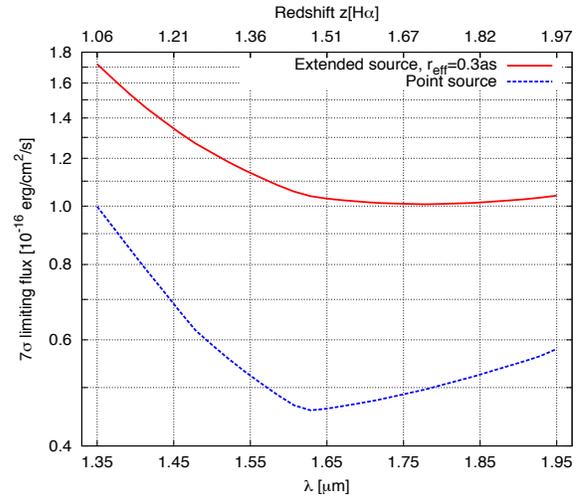

**Figure 3:** Emission line sensitivity of the WFIRST-2.4 high-latitude spectroscopic survey. The blue curve shows 7$\sigma$ point source sensitivities, and the red curve shows extended source ($r_{eff}$ = 0.3 arcsec, exponential profile) sensitivities. The depth is observed-frame (not corrected for Galactic extinction). The depth at a particular point on the sky depends on the zodiacal light level and the number of dithered exposures that cover the location without chip gaps or cosmic rays, but most of the HLS area will be observed to at least the depth shown here.

The high-cadence bulge survey, the high-latitude imaging and spectroscopic surveys, and the synoptic, deep supernova survey will all support a wide range of science investigations beyond dark energy and planet discovery. These would be supported by a Guest Investigator program, analogous to archival programs on HST and the other Great Observatories. Investigations requiring new observations would come under the Guest Observer program.

The anticipated schedule and cost of WFIRST-2.4 are discussed in detail in the SDT report. The development phase (B/C/D) from mission start through launch is 79 months, including 7 months of reserve. Consumables on the spacecraft would be budgeted for a 10-year mission life. WFIRST-2.4 is a remarkable facility, a Hubble-quality telescope equipped with extremely powerful instruments, so there will be strong science drivers for an extended mission.

| WFIRST-2.4 Design Reference Mission Capabilities | | | | | | |
|---|---|---|---|---|---|---|
| Imaging Capability | 0.281 deg$^2$ | | 0.11 arcsec/pix | | 0.6 – 2.0 μm | |
| Filters | Z087 | Y106 | J129 | H158 | F184 | W149 |
| Wavelength (μm) | 0.760-0.977 | 0.927-1.192 | 1.131-1.454 | 1.380-1.774 | 1.683-2.000 | 0.927-2.000 |
| PSF EE50 (arcsec) | 0.11 | 0.12 | 0.12 | 0.14 | 0.14 | 0.13 |
| Spectroscopic Capability | Grism (0.281 deg$^2$) 1.35 – 1.95 μm, R = 550-800 | | | IFU (3.00 x 3.15 arcsec) 0.6 – 2.0 μm, R = ~100 | | |

| Baseline Survey Characteristics | | | | | |
|---|---|---|---|---|---|
| Survey | Bandpass | Area (deg$^2$) | Depth | Duration | Cadence |
| Exoplanet Microlensing | Z, W | 2.81 | n/a | 6 x 72 days | W: 15 min Z: 12 hrs |
| HLS Imaging | Y, J, H, F184 | 2000 | Y = 26.7, J = 26.9 H = 26.7, F184 = 26.2 | 1.3 years | n/a |
| HLS Spectroscopy | 1.35 – 1.95 μm | 2000 | 0.5x10$^{-16}$ erg/s/cm$^2$ @ 1.65 μm | 0.6 years | n/a |
| SN Survey | | | | 0.5 years (in a 2-yr interval) | 5 days |
| Wide | Y, J | 27.44 | Y = 27.1, J = 27.5 | | |
| Medium | J, H | 8.96 | J = 27.6, H = 28.1 | | |
| Deep | J, H | 5.04 | J = 29.3, H = 29.4 | | |
| IFU Spec | 7 exposures with S/N=3/pix, 1 near peak with S/N=10/pix, 1 post-SN reference with S/N=6/pix Parallel imaging during deep tier IFU spectroscopy: Z, Y, J, H ~29.5, F184 ~29.0 | | | | |

| Guest Observer Capabilities | | | | | | |
|---|---|---|---|---|---|---|
| 1.4 years of the 5 year prime mission | | | | | | |
| | Z087 | Y106 | J129 | H158 | F184 | W149 |
| Imaging depth in 1000 seconds (m$_{AB}$) | 27.15 | 27.13 | 27.14 | 27.12 | 26.15 | 27.67 |
| t$_{exp}$ for σ$_{read}$ = σ$_{sky}$ (secs) | 200 | 190 | 180 | 180 | 240 | 90 |
| Grism depth in 1000 sec | S/N=10 per R=~600 element at AB=20.4 (1.45 μm) or 20.5 (1.75 μm) t$_{exp}$ for σ$_{read}$ = σ$_{sky}$: 170 secs | | | | | |
| IFU depth in 1000 sec | S/N=10 per R~100 element at AB=24.2 (1.5 μm) | | | | | |
| Slew and settle time | chip gap step: 13 sec, full field step: 61 sec, 10 deg step: 178 sec | | | | | |

| Optional Coronagraph Capabilities | |
|---|---|
| 1 year in addition to the 5-year primary mission, interspersed, for a 6-year total mission | |
| Field of view | Annular region around star, with 0.2 to 2.0 arcsec inner and outer radii |
| Sensitivity | Able to detect gas-giant planets and bright debris disks at the 1 ppb brightness level |
| Wavelength range | 400 to 1000 nm |
| Image mode | Images of full annular region with sequential 10% bandpass filters |
| Spectroscopy mode | Spectra of full annular region with spectral resolution of 70 |
| Polarization mode | Imaging in 10% filters with full Stokes polarization |
| Stretch goals | 0.1 arcsec inner annulus radius, and super-Earth planets |

**Table 1**: WFIRST-2.4 design reference mission observing program. The quoted magnitude/flux limits are for point sources, 5σ for imaging, 7σ for HLS spectroscopy.

## 3 WFIRST-2.4 SCIENCE PROGRAMS

### 3.1 Dark Energy

The accelerating expansion of the universe is the most surprising cosmological discovery in many decades, with profound consequences for our understanding of fundamental physics and the mechanisms that govern the evolution of the cosmos. The two top-level questions of the field are:

1. Is cosmic acceleration caused by a new energy component or by the breakdown of General Relativity (GR) on cosmological scales?
2. If the cause is a new energy component, is its energy density constant in space and time, or has it evolved over the history of the universe?

A constant energy component, a.k.a. a "cosmological constant," could arise from the gravitational effects of the quantum vacuum. An evolving energy component would imply a new type of dynamical field. Gravitational explanations could come from changing the action in Einstein's GR equation, or from still more radical modifications such as extra spatial dimensions. Observationally, the route to addressing these questions is to measure the histories of cosmic expansion and growth of structure with the greatest achievable precision over the widest accessible redshift range.

Figure 4 presents an overview of the WFIRST-2.4 dark energy program. Type Ia SNe will be used to measure "standard candle" distances out to $z = 1.7$, calibrated against a (ground-based) sample observed in the local Hubble flow. With the model of statistical and systematic errors detailed in the SDT report, the aggregate precision of these measurements is 0.20% at z < 1 (error-weighted <z> = 0.50) and 0.34% at z > 1 (<z> = 1.32). The baryon acoustic oscillation (BAO) feature in galaxy clustering provides a "standard ruler" for distance measurement, calibrated in absolute units, independent of $H_0$. The galaxy redshift survey (GRS) enables measurements of the angular diameter distance $D_A(z)$ and the expansion rate H(z) using Hα emission line galaxies at z = 1 - 2 and [OIII] emission line galaxies at z = 2 – 3, with aggregate precision ranging from 0.40% to 1.8% (see figure). The imaging survey will enable measurements of dark matter clustering via cosmic shear and via the abundance of galaxy clusters with mean mass profiles calibrated by weak lensing; we expect 40,000 M ≥ $10^{14} M_{sun}$ clusters in the 2000 $deg^2$ area of the high-latitude survey. These data constrain the amplitude of matter fluctuations at 0 < z < 2 and provide additional leverage on the redshift-distance relation. The expected aggregate precision on the fluctuation amplitude as an isolated parameter change is ≈ 0.15% at z < 1 and 0.3-0.5% at z > 1. Redshift-space distortions in the GRS provide an entirely independent approach to measuring the growth of structure, with aggregate precision ≈ 1% at z = 1-2.

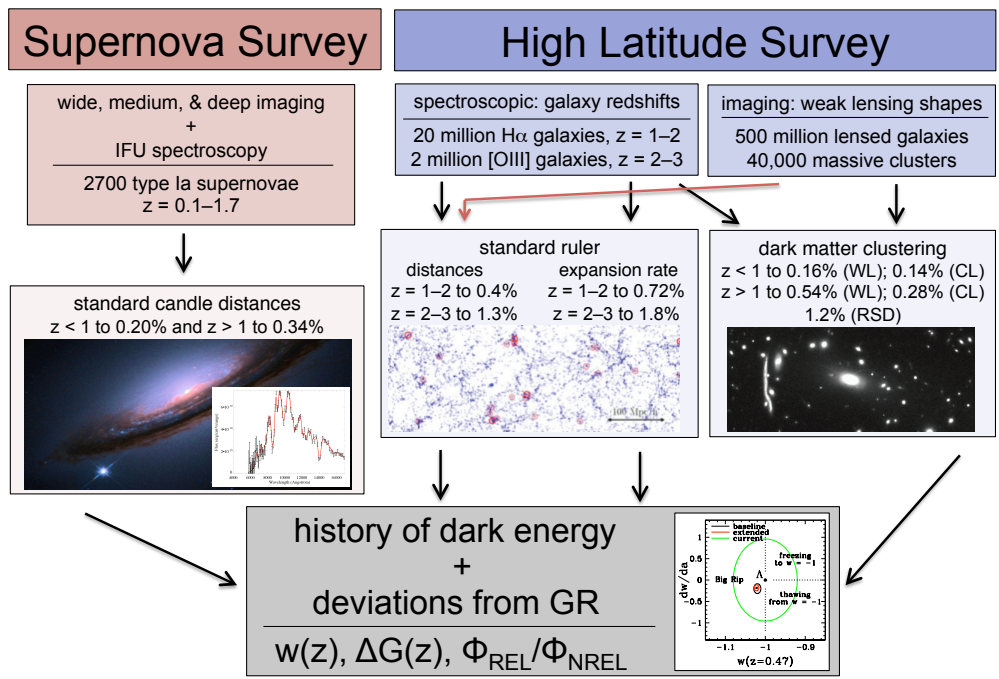

**Figure 4:** A high-level view of the WFIRST-2.4 dark energy program. The supernova (SN) survey will measure the cosmic expansion history through precise spectrophotometric measurements of more than 2700 supernovae out to redshift $z = 1.7$. The high-latitude survey (HLS) will measure redshifts of more than 20 million emission-line galaxies and shapes (in multiple filters) of more than 500 million galaxies. The former allow measurements of "standard ruler" distances through characteristic scales imprinted in the galaxy clustering pattern, while the latter allow measurements of matter clustering through the "cosmic shear" produced by weak gravitational lensing and through the abundance of galaxy clusters with masses calibrated by weak lensing. As indicated by crossing arrows, weak lensing measurements also constrain distances, while the galaxy redshift survey provides an alternative measure of structure growth through the distortion of redshift-space clustering induced by galaxy motions. Boxes in the middle layer list the forecast aggregate precision of these measurements in different ranges of redshift. These high-precision measurements of multiple cosmological observables spanning most of the history of the universe lead to stringent tests of theories for the origin of cosmic acceleration, through constraints on the dark energy equation-of-state parameter $w(z)$, on deviations $\Delta G(z)$ from the growth of structure predicted by General Relativity, or on deviations between the gravitational potentials that govern relativistic particles (and thus weak lensing) and non-relativistic tracers (and thus galaxy motions).

These high-precision measurements over a wide range of redshifts in turn lead to powerful constraints on theories of cosmic acceleration. If the cause of acceleration is a new energy component, then the key physical characteristic is the history w(z) of the equation-of-state parameter w = P/ε, the ratio of pressure to energy density. A cosmological constant has w = -1 at all times, while dynamical dark energy models have w ≠ -1 and an evolutionary history that depends on the underlying physics of the dark energy field. If the cause of acceleration is a breakdown of GR on cosmological scales, then it may be detected in a deviation between the measured growth history G(z) and the growth predicted by GR given the measured expansion history. Alternatively, some modified gravity theories predict a mismatch between the gravitational potential inferred from weak lensing (in Figure 4, cosmic shear and clusters) and the gravitational potential that affects motions of non-relativistic tracers, which governs redshift-space distortions in the galaxy redshift survey. Figure 5 illustrates the forecast precision of WFIRST-2.4 constraints on parameters of dark energy, in a model where w(z) is a linear function of expansion parameter a = $(1+z)^{-1}$. These constraints would be a dramatic improvement on current knowledge, enabling robust discovery of devia-tions from a cosmological constant that are within the errors of existing measurements.

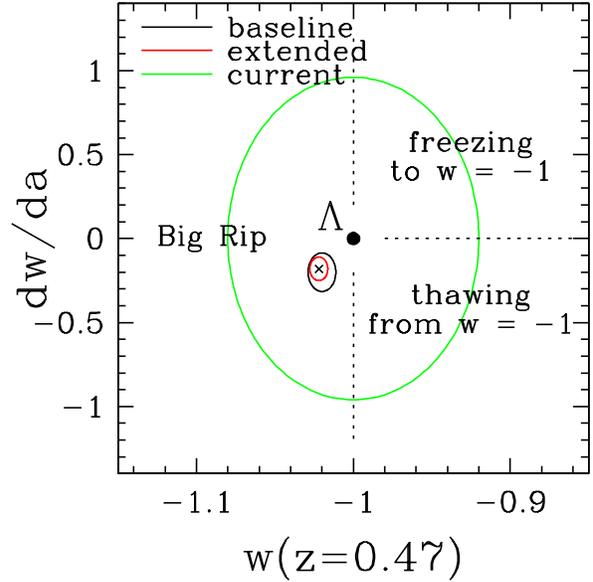

**Figure 5:** Forecast constraints on dark energy parameters from WFIRST-2.4 compared to current knowledge, for a model in which the dark energy equation-of-state parameter is w(z) = $w_0$ + $w_a$(1-a) with a = $(1+z)^{-1}$. Ellipses show Δ$\chi^2$ = 1 error contours on the value of w at redshift z = 0.47 (the redshift at which it is best determined by WFIRST-2.4) and its derivative dw/da = -$w_a$. The green ellipse, centered here on the cosmological constant model (w = -1, dw/da = 0), represents current state-of-the-art constraints from a combination of CMB, SN, BAO, and $H_0$ data, based on Anderson et al. (2012). For this figure, we have imagined that the true cosmology is w(z=0.47) = -1.022 and dw/da = -0.18, well within current observational constraints. The black ellipse shows the error forecast for the baseline WFIRST-2.4 supernova, galaxy redshift, and weak lensing surveys, combined with CMB data from Planck, a local supernova calibrator sample, and measurements of BAO and redshift-space distortions from the SDSS-III BOSS survey at z < 0.7. The red ellipse shows the "extended" case in which the precision of the WFIRST-2.4 measurements (but not the Planck, local SN, or BOSS measurements) is increased by a factor of two, as a result of a longer observing program in an extended mission, better control of systematic uncertainties, or both. Legends indicate physically distinct regions of the parameter space: a cosmological constant (Λ), scalar field models that are "freezing" towards or "thawing" from w = -1, and models with w < -1 (often referred to as "phantom energy") in which increasing acceleration leads to a "big rip" at a finite time in the future.

The high statistical precision of the WFIRST-2.4 measurements places stringent demands on the control of systematic biases, and the mission is designed with this point foremost in mind. For SNe, the use of a stable, space-based observing platform and near-IR measure-ments already mitigates key systematics affecting ground-based surveys. The IFU employed on WFIRST-2.4 enables flux measurements in synthetic redshifted bands that avoid the need for k-corrections, better separation of supernova and host-galaxy light, more accurate flux calibration, and the use of spectroscopic indicators to mitigate evolutionary effects. For weak lensing, the HLS reaches effective source densities of 54, 61, and 44 galaxies arcmin$^{-2}$ in the J, H, and F184 filters, respectively, enabling cross-checks among three shear auto-correlation functions and three cross-filter correlation functions. Each weak lensing galaxy appears in 5 – 8 separate exposures at two different roll angles. The high level of redundancy allows detailed tests of the long chain of calibrations and assumptions that go into any weak lensing analysis. For the galaxy redshift survey, the high space density reached by WFIRST-2.4 will make it possible to measure higher order clustering statistics and split the data into subsamples to test models of galaxy bias, which are the primary source of uncertainty in deriving cosmological constraints from redshift-space distortions and intermediate-scale clustering.

## 3.2 Astrophysics with the High-Latitude Surveys

While the dark energy program drives the design of the high-latitude imaging and spectroscopic surveys, the scientific motivation for them is much broader. These surveys will provide an extraordinary resource for a wide range of investigations. The SDT report discusses a few of these topics as examples, though the list is far from exhaustive:

- *The First Billion Years of Cosmic History*
  Deep, multi-band, near-IR imaging over 2000 deg$^2$ will be a revolutionary data set for understanding the first generations of galaxies and quasars. Extrapolations of high-redshift luminosity functions are uncertain, but they suggest that the HLS will detect ~ 30,000 $z > 8$ galaxies above the 10$\sigma$, mag ≈ 26 limit, and ~ 100 at $z > 10$ (rising to ~ 1000 at the 5$\sigma$ limit). Some of these galaxies will be highly magnified by the gravitational lensing of foreground clusters, making them ideal targets for follow-up observations with JWST. The synergy of a wide-field telescope that can discover the most luminous or highly magnified systems and a large aperture telescope that can characterize them in detail is essential for understanding the earliest galaxies.

- *Mapping Dark Matter on Intermediate and Large Scales*
  The HLS will produce weak lensing catalogs with surface densities of ~ 75 galaxies arcmin$^{-2}$ in co-added J and H bands. These can be used to map the distributions of dark matter in clusters and superclusters with much higher fidelity than achievable from the ground (or from Euclid, which has less than half this source density). Weak+strong lensing measurements of cluster mass profiles and substructure, compared to predictions from numerical simulations, will yield new insights into the mechanisms of structure formation and the interactions between baryons and dark matter. Merging/colliding clusters, such as the famed Bullet Cluster, are especially interesting systems, allowing novel tests of the properties of dark matter and the impact of cluster assembly on galaxy-scale halos and intracluster gas. Deeper observations in the Guest Observer program could achieve surface densities of 200 – 300 arcmin$^{-2}$ for detailed study of interesting systems, with a survey speed fast enough to yield statistically informative samples.

- *Kinematics of Stellar Streams in Our Local Group of Galaxies*
  Using the high density of slightly resolved, medium-brightness galaxies as references, WFIRST-2.4 will measure accurate proper motions to a limit several magnitudes fainter than the GAIA mission. For G stars at V ≈ 20, we estimate accuracy of about 125 µas/year from the 2-year baseline of the HLS, which could be improved to 50 µas/year in GO programs that extend the baseline to 5 years. This depth probes the main sequence turnoff to distances of ~ 10 kpc and red giants throughout the Galactic halo. Tidal streams from disrupted dwarfs or star clusters have velocity dispersions of ~ 10 km s$^{-1}$, so with proper motions of this accuracy they will stand out at high contrast from the background halo stars, which have a dispersion of ~ 150 km s$^{-1}$. Measurements of the kinematics and structure of cold tidal streams can determine the structure of the Milky Way's gravitational potential and detect or rule out the small scale perturbations from dark matter subhalos that are a critical prediction of the cold dark matter scenario.

- *Discovering the Most Extreme Star Forming Galaxies and Quasars*
  The HLS spectroscopic survey will enable the largest census yet performed of powerful emission-line galaxies and quasars up to (and possibly beyond) lookback times of 90% of the current age of the universe. Samples of H$\alpha$, [OIII], and [OII] emitters will measure the contribution of luminous galaxies to the global star formation rate at $1 < z < 4$, when most stars in the universe formed. At redshifts $8 < z < 15$ (roughly 200 – 600 million years after the Big Bang), Lyman-alpha emitting galaxies will be detectable if they have attenuated star formation rates of at least 100 – 200 solar masses per year. The survey will detect ~ 2600 z > 7 quasars, with an estimated 20% of them at z > 8, probing the assembly of billion solar mass black holes in the first Gyr of cosmic history and finding the rare, bright backlights that JWST and ground-based telescopes can use to trace the earliest enrichment of the intergalactic medium with heavy elements (e.g., via CIV absorption).

## 3.3 Exoplanet Science

The discovery of planetary companions to Sun-like stars was one of the greatest astronomical breakthroughs of recent decades, exciting both the scientific community and the broader public. Nature has surprised astronomers with the enormous and unexpected diversity of exoplanetary systems, containing planets with physical properties and orbital architectures that are radically different from our own Solar System. WFIRST-2.4 will advance our understanding of exoplanets along two complementary fronts: the statistical approach of determining the demographics of exoplanetary systems over broad regions of parameter space and the detailed approach of characterizing (with coronagraphy) the properties of a small number of nearby exoplanets.

### 3.3.1 *The Demographics of Exoplanets*

The WFIRST-2.4 microlensing survey will continuously monitor a total of 2.81 deg$^2$ in the Galactic bulge for six 72-day campaigns, with a cadence of 15 minutes in a wide filter (0.927 – 2.00 µm) for planet discovery and 12 hours in Z-band for characterization of source and lens stars. This survey will detect tens of thousands of stellar microlensing events, and in cases where the lensing star has an orbiting planet it can (when the alignment is favorable) produce a distinctive perturbation to the light curve. For a large fraction of these events, WFIRST-2.4 will be able to completely characterize the lensing system, including the host star and planet mass, the planet orbital separation in physical units, and the distance to the system. It will do this through a combination of subtle deviations from the simplest light curve shapes, including finite-source effects, "parallax" caused by motion of the earth and the observatory's geosynchronous orbit, and source astrometric motion induced by the host lens, together with the isolation and direct measurement of the light from luminous host lenses afforded by the high angular resolution. Figure 6 provides two examples of simulated

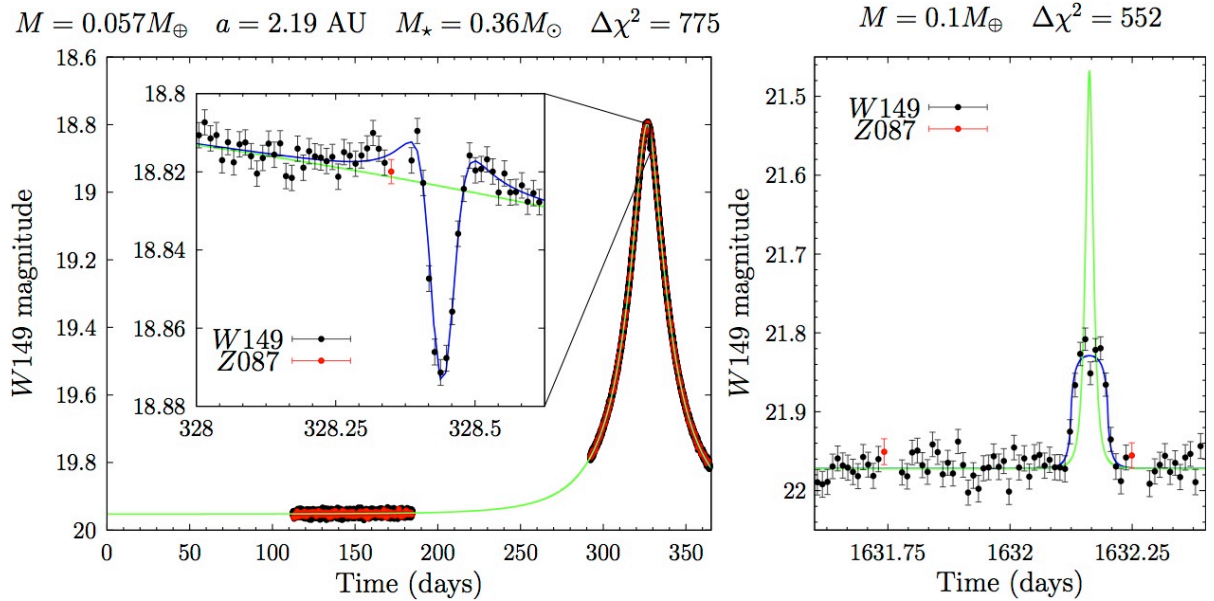

**Figure 6:** Examples of simulated WFIRST-2.4 microlensing event light curves with detected planetary signals. On the left, the full panel shows the overall shape of the microlensing light curve, with points marking observations from two different bulge observing seasons (wide filter in black, Z filter in red) and the green curve showing the best-fit point-microlens model. The inset panel zooms in on a 16-hour interval where a deviation reveals the presence of a Mercury-mass planet orbiting a 0.36 solar mass star with a semi-major axis of 2.19 AU. The right panel shows a simulated detection of a free-floating Mars-mass planet, where the short duration already indicates a low mass lens and the flat-topped (instead of sharply peaked) light curve shows that Einstein ring of the lens magnifies only a fraction of the source star's surface.

WFIRST-2.4 microlensing events, detecting a Mercury-mass planet orbiting an M-dwarf and a free-floating Mars-mass planet.

The Kepler satellite has made a crucial contribution to our understanding of planet formation by systematically assaying the population of "hot" and "warm" planets down to near-earth radii. As illustrated in Figure 7, WFIRST-2.4 is an ideal complement to Kepler because it is most sensitive to "cold" planets at distances of a few AU from the parent star. The sensitivity of the WFIRST-2.4 microlensing survey extends to masses of Mars and below. Crucially, microlensing is also sensitive to free-floating planets ejected from the systems where they were born, which many planet-formation theories predict to be as common as bound planets. One of the central questions in exoplanet astronomy is the frequency of "habitable" worlds that could have liquid water at their surfaces. Kepler and WFIRST-2.4 address this question in complementary ways, approaching the habitable zone from the inside-out and the outside-in, respectively. The sensitivity of microlensing is also complementary to that of ground-based radial velocity and transit surveys.

Forecasts of planet yields are necessarily uncertain, because they depend on the very questions about exoplanet demographics that WFIRST-2.4 seeks to answer. We can nonetheless make predictions based on plausible extrapolations of current knowledge (as described in Appendix D of the SDT report). These forecasts suggest that WFIRST-2.4 will detect about 3000 planets overall, including 1000 "super-earths" (roughly 10 times the mass of earth), 300 earth-mass planets, and 40 Mars-mass planets. These detections would enable measurement of the mass function of cold exoplanets to better than ~10% per decade in mass for masses > 0.3 $M_{earth}$, and an estimate of the frequency of Mars-mass embryos accurate to ~15%. If there is one free-floating earth-mass planet per star in the Galaxy, then WFIRST-2.4 will detect about 40 of them, and much larger numbers of more massive free-floating planets. Forecast yields for WFIRST-2.4 are noticeably higher (about 25%) than those for DRM1, but the more important gain is the larger fraction of events that can be fully characterized thanks to higher angular resolution and the parallax effect in geosynchronous orbit.

### 3.3.2 Characterization of Exoplanets and Debris Disks

Our understanding of the internal structure, atmospheres, and evolution of planets was originally developed through models that were tuned to explain the detailed properties of the planets in our own solar system.

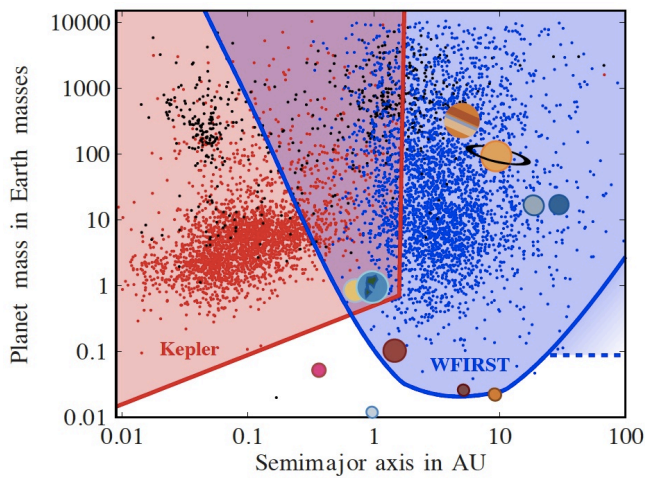

**Figure 7:** The complementary sensitivity of WFIRST-2.4 (blue) and Kepler (red) to planets in different ranges of mass and semi-major axis. Kepler is sensitive to the abundant, hot and warm terrestrial planets with separations less than about 1.5 AU. WFIRST-2.4 is sensitive to Earth-mass planets with separations greater than 1 AU, as well as planets down to roughly twice the mass of the moon at slightly larger separations. WFIRST-2.4 is also sensitive to unbound planets with masses as low as Mars. The small red points show candidate planets from Kepler, whereas the small blue points show simulated detections by WFIRST-2.4. The solar system planets are also shown, as well as the Moon, Ganymede, and Titan. The expected number of WFIRST-2.4 detections is large, with roughly 2800 bound and hundreds of free-floating planet discoveries. WFIRST-2.4 and Kepler together cover the entire planet discovery space in mass and orbital separation, providing the comprehensive survey of exoplanet demographics necessary to fully understand the formation and evolution of planetary systems. The large area of WFIRST-2.4 discovery space combined with the large number of detections essentially guarantees a number of unexpected and surprising discoveries.

Surveys of exoplanetary systems have led to the realization that there exists a diversity of worlds with very different properties and environments than those in our solar system. The best hope of understanding the physical properties of this diversity of worlds is through detailed characterization of the properties of individual planets and their atmospheres. High-contrast, high-angular-resolution direct imaging provides the critical approach to studying the detailed properties of exoplanets. Images and spectra of directly imaged planets provide some of the most powerful diagnostic information about the structure, composition, and physics of planetary atmospheres, which in turn can provide constraints on the origin and evolution of these systems. The direct imaging technique is also naturally applicable to the nearest and brightest, and thus best-characterized, solar systems. High contrast imaging is also ideally suited to studying the diversity and properties of debris disks around the nearest stars; these disks serve as both fossil records of planet formation, and signposts of extant planets through their dynamical influences.

With a contrast level of $10^{-9}$ and an inner working angle of less than 0.2", WFIRST-2.4 equipped with a coronagraph will perform a survey of up to 200 of the nearest stars (see Figure 8). This survey will directly image over a dozen known radial velocity planets and will discover an additional dozen previously unknown ice and gas giants. The majority of these planets will also be characterized using R~70 spectra in the wavelength range 400-1000 nm via an integral field spectrograph. These spectra will allow the detection of features expected due to methane, water, and alkali metals, reveal the signature of Rayleigh scattering, and easily distinguish between different classes of planets (i.e., Neptunes versus Jupiters), and planets with very different metallicities.

This survey will also be sensitive to debris disks with a few times the solar system's level of dust in the habitable zones and asteroid belts of nearby (~10 pc) sun-like stars. The high sensitivity and spatial resolution (0.05 arcsec is 0.5 AU at 10 pc) of WFIRST-2.4 images will map the large-scale structure of these disks, revealing asymmetries, asteroid belts, and gaps due to unseen planets. WFIRST-2.4 will make the most sensitive measurements yet of the amount of dust in or near the habitable zones of nearby stars. This is important for assessing the difficulty of imaging Earth-like planets with future missions as well as for understanding nearby planetary systems. Finally, spectrophotometry of these disks at the full range of available wavelengths from 400-1000 nm provides constraints on dust grain size

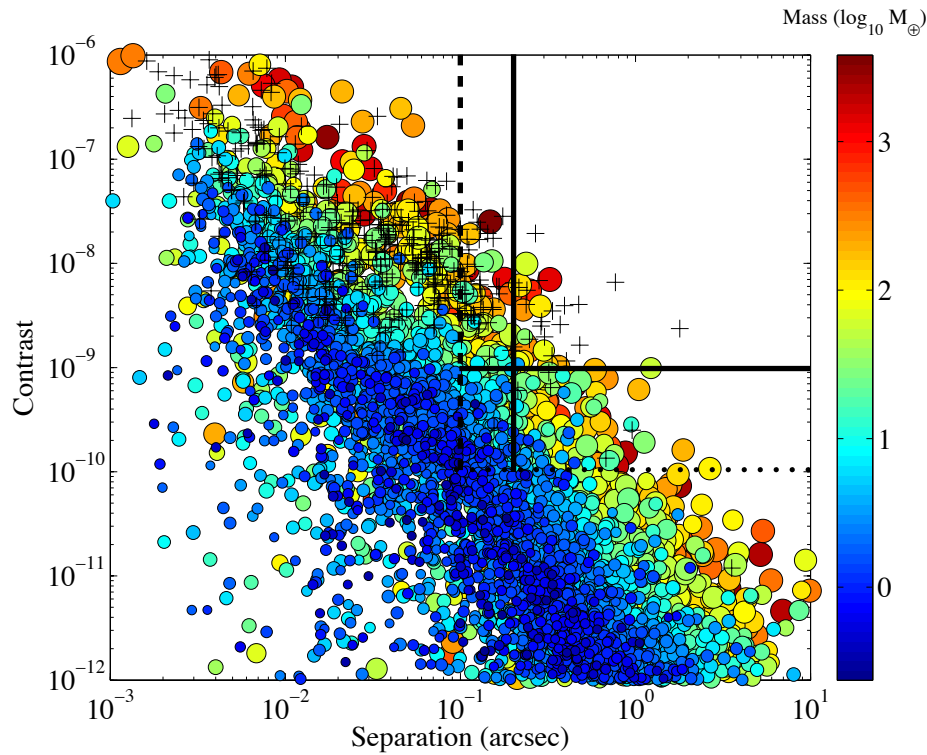

**Figure 8: Sensitivity of the WFIRST-2.4 coronagraph for imaging planets around nearby stars.** Solid black lines mark the baseline technical goal of 1 ppb contrast and 0.2 arcsec inner working angle, while the dotted lines show the more aggressive goals of 0.1 ppb and 0.1 arcsec. Colored circles show a snapshot in time from a simulation of model planets, ranging in size from Mars-like to several times the radius of Jupiter, placed in orbit around ~200 of the nearest stars within 30 pc. The model assumes roughly four planets per star with a mixture of gas giants, ice giants, and rocky planets, and a size and radius distribution consistent with Kepler results and extrapolated to larger semi-major axis and lower mass. Color indicates planet mass while size indicates planet radius. Crosses represent known radial velocity planets. At the baseline sensitivity, WFIRST-2.4 would be able to image a number of known radial velocity planets, and it would detect a number of new gas and ice giants. If the more aggressive goals are achievable, then the number of detectable planets increases significantly, including a small number of water planets and Super-Earths.

and composition.

If an enhanced coronagraph capable of achieving contrast levels of $10^{-10}$ and inner working angles of ~0.1" proves possible, WFIRST-2.4 might even be capable of detecting rocky, Super-Earth-like, planets around a small number of nearby stars. Adding polarimetric imaging capability would make it easier to discriminate dust clumps from planets as well as characterize the properties of planetary atmospheres and dust grains in disks.

Advancing the technology for direct imaging of exoplanets was the top priority medium-scale space investment recommended by *NWNH*. Developing a coronagraph with active wavefront control for WFIRST-2.4 accomplishes this objective and, thanks to the 2.4m telescope, achieves far more real science than would be possible on a technology demonstration mission with a much smaller aperture. Coronography on WFIRST-2.4 would be a major step towards the long-term goal of a mission that can image habitable earth-mass planets around nearby stars and measure their spectra for signs of life.

### 3.4 The Guest Observer Program

The combination of a 0.28 deg$^2$ field with the near-IR sensitivity and angular resolution afforded by a 2.4m telescope gives WFIRST-2.4 extraordinary discovery potential. In the 5-year prime mission, 1.4 years would be allocated to GO programs. If WFIRST-2.4 enters an extended mission phase, observing time would be competed, with the balance between large and small programs to be decided by the Time Allocation Committee. This will afford a large number of unique, high impact projects from characterizing nearby brown dwarfs to probing the epoch of reionization.

Figure 9 illustrates the power of WFIRST-2.4 for studies of resolved stellar populations and extended low surface brightness structure around nearby galaxies. The left panel shows a near-IR color-magnitude diagram of the globular cluster 47 Tucanae from a 3-orbit integration with HST/WFC3/IR (Kalirai et al. 2012). The sharp "kink" on the lower main sequence is caused by collisionally induced absorption of $H_2$. In an uncrowded field, a 10 ksec exposure with WFIRST-2.4 can reach the *bottom* of the hydrogen burning main sequence at a distance of 30 kpc, and the main sequence turnoff of an old stellar population out to 500 kpc. Red giant clump stars are detectable at 2 Mpc, and the tip of the red giant branch at 15 Mpc (all numbers $10\sigma$). The depth is reduced in crowded fields or against the background light of a galaxy disk or spheroid, but in such cases the high angular resolution of WFIRST-2.4 gives it an even larger advantage over near-IR observations from ground-based telescopes or Euclid. The right panel, from Martinez-Delgado et al. (2010), shows the extended low surface brightness structure around the galaxy NGC 5055, revealing the long-lived signatures of hierarchical galaxy assembly beyond the traditional "optical radius." The WFIRST-2.4 HLS will probe this low surface brightness structure around many thousands of nearby galaxies, and GO observations can extend these investigations to greater depth and to systematic coverage of different environments, from the centers of voids to the core of the Virgo cluster. For galaxies within a few Mpc, WFIRST-2.4 will resolve individual red giant stars as well as imaging the low surface brightness background of main sequence stars. These investigations will show how recent accretion and minor mergers affect the growth of galaxies across a wide spectrum of mass, morphology, and environment.

Appendix A of the SDT report contains ~ 50 potential GO or Guest Investigator science programs, each described in one page, contributed by members of the broader astronomical community. Here we simply list their titles to illustrate the remarkable range of ideas for scientific investigations enabled by WFIRST-2.4:

***Planetary Bodies***
- *A Full Portrait of the Kuiper Belt, Including Size Distributions, Colors, and Bimodality*
- *The Outer Solar System from Neptune to the Oort Cloud*
- *Free-floating Planets in the Solar Neighborhood*
- *Measuring Planet Masses with Transit Timing Variations*
- *Exoplanet Spectroscopy with WFIRST*
- *WFIRST: Additional Planet Finding Capabilities – Astrometry*
- *WFIRST: Additional Planet Finding Capabilities – Transits*

***Stellar Astrophysics***
- *Stellar and Substellar Populations in Galactic Star Forming Regions*
- *Identifying the Coldest Brown Dwarfs*
- *Stellar Fossils in the Milky Way*
- *The Infrared Color-Magnitude Relation*
- *Finding the Closest Young Stars*
- *The Most Distant Star-Forming Regions in the Milky Way*
- *Super-resolution Imaging of Low-mass Stars with Kernel-phase and Precision Wavefront Calibration with Eigen-phase*

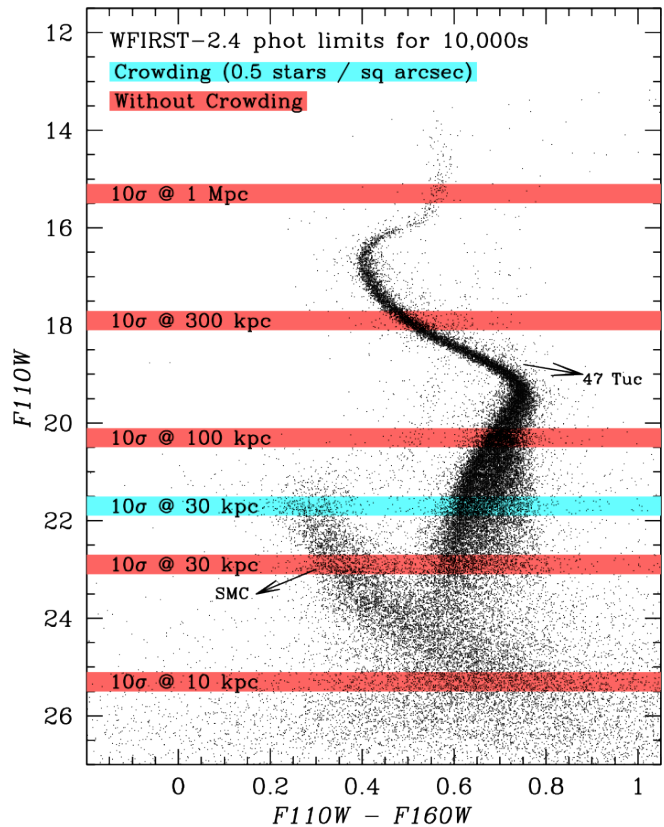

Figure 9: The power of WFIRST-2.4 for GO programs in the Milky Way and local volume. The left panel shows the near-IR color-magnitude diagram (roughly J vs. J-H, Vega-based magnitudes) from a 3-orbit HST WFC3 observation of the globular cluster 47 Tucanae, from Kalirai et al. (2012). The distinctive kink on the lower main sequence marks the onset of $H_2$ opacity, and it provides a valuable diagnostic for breaking degeneracies among distance, age, reddening, and metallicity in star cluster studies. Horizontal bars mark the approximate $10\sigma$ point source depth of a 10 ksec WFIRST-2.4 observation at different distances; red bars are for uncrowded fields, while the cyan bar shows a globular cluster-like crowding level at 30 kpc. Individual main sequence stars can be detected to 500 kpc. Red giant clump stars (not marked) could be detected to about 2 Mpc, while the brightest magnitude of the red giant branch could be detected to ~ 15 Mpc. The right panel, based on Martínez-Delgado et al. (2010), superposes the WFIRST-2.4 field of view on a deep ground-based image of the galaxy NGC 5055 (Messier 63), revealing low surface brightness signatures of dynamical disturbances far beyond the traditional "optical radius" of the galaxy disk (colored inset). WFIRST-2.4 will probe low surface brightness structure around thousands of nearby galaxies, and at the 12 Mpc distance of NGC 5055 a 10 ksec GO observation would readily detect individual stars near the tip of the red giant branch.

- *Detecting and Characterizing Neutron Stars, Black Holes with Astrometric Microlensing*

**Galactic Astrophysics and the Local Volume**

- *Proper Motions and Parallaxes of Disk and Bulge Stars*
- *Quasars as a Reference Frame for Proper Motion Studies*
- *The Detection of the Elusive Stellar Counterpart of the Magellanic Stream*
- *Near-field Cosmology: Finding the Faintest Milky Way Satellites*
- *The Mass of the Milky Way*
- *Distinguishing Between Cold and Warm Dark Matter with WFIRST*
- *Finding (or Losing) Those Missing Satellites*
- *Mapping the Potential of the Milky Way with Tidal Debris*
- *Dissecting Nearby Galaxies*
- *Galaxy Evolution from Resolved Stellar Pops: Halo Age Distributions of the Local Volume*
- *Substructure Around Galaxies Within 50 Mpc*
- *Resolved Stellar Populations in Nearby Galaxies*
- *Deep Surface Photometry of Galaxies and Galaxy Clusters*

*Extragalactic Astrophysics*
- Galaxy Structure and Morphology
- Strong Lensing
- Searching for Extreme Shock-dominated Galaxy Systems from z = 1 – 2
- Mapping the Distribution of Matter in Galaxy Clusters
- Merging Clusters of Galaxies
- Group-Scale Lenses: Unexplored Territory
- The Evolution of Massive Galaxies: The Formation and Morphologies of Red Sequence Galaxies
- Finding and Weighing Distant, High Mass Clusters of Galaxies
- Probing the Epoch of Reionization with Lyman-Alpha Emitters
- Obscured Quasars
- The Faint End of the Quasar Luminosity Function
- Strongly Lensed Quasars
- High-Redshift Quasars and Reionization
- Characterizing the Sources Responsible for Reionization
- Finding the First Cosmic Explosions with WFIRST
- Resolved Stellar Population Studies in z ~ 2 Star Forming Galaxies

*General*
- Synergy Between LSST and WFIRST
- Synergies Between Euclid and WFIRST
- The Shapes of Galaxy Haloes from Gravitational Flexion
- WFIRST and IRSA: Synergy Between All-Sky IR Surveys
- Near Infrared Counterparts of Elusive Binary Neutron Star Mergers

A separate Appendix of the SDT report is devoted to synergistic science programs involving both WFIRST-2.4 and JWST, on topics that range from AGB stars and dust production to finding the first stars via pair instability supernovae. The power of these combined programs provides strong motivation for launching WFIRST-2.4 early in the next decade.

## 4 CONCLUDING REMARKS

As emphasized by *NWNH*, a wide-field infrared telescope in space addresses some of the most pressing issues in contemporary astrophysics and opens an enormous frontier for new discoveries. Such a telescope would be highly complementary to LSST, to Euclid, to JWST, and to the next generation of giant ground-based observatories. The availability of a 2.4m telescope makes this prospect even more attractive, with greater sensitivity and angular resolution, and the potential for sensitive coronagraphic observations of planets around nearby stars. Relative to the 1.3m DRM1 design, WFIRST-2.4 is a more powerful mission with comparable cost and lower risk. JWST and WFIRST-2.4 will be a formidable complement of Great Observatories for the 2020s.